\documentclass{ws-p8-50x6-00}

\begin{document}

\def\a{{a}}
\def\b{{b}}
\def\m{{m}}

\title{The $7/11$ Rule and an Estimate of $\m_\rho$}

\author{Maarten Golterman$^{\a 1}$ and Santi Peris$^\b$}

\address{{}$^a$Department of Physics, Washington University, St. Louis MO
63130, USA\\E-mail: maarten@aapje.wustl.edu\\
{}$^b$IFAE, Univ. Autonoma de Barcelona, Bellaterra, 08193 Barcelona, Spain
\\E-mail: peris@ifae.es}

\maketitle

The two-point functions for the vector and axial currents
$V_\mu=\overline{d}\gamma_\mu u$ and $A_\mu=\overline{d}\gamma_\mu
\gamma_5 u$ contain information on the hadronic resonances
$\rho$, $\rho'$, $a_1$, {\it etc.}  While it has thus far not
been possible to calculate these two-point functions from QCD,
their behavior is constrained at low energy by 
Chiral Perturbation Theory (ChPT),  at high energy by the OPE, and at all
scales by large $N_c$, in which limit
the two-point functions are saturated by an infinite tower of
narrow resonances:
\begin{equation}
\frac{1}{\pi}{\rm Im}\,\Pi_V(t)=\sum_{n=1}^\infty 2f_n^2m_n^2
\delta(t-m_n^2)\,,\label{RESON}
\end{equation}
with $m_1=m_\rho$, $f_1=f_\rho$ the EM decay constant of the $\rho$.
Similarly, the axial two-point function is a tower of resonances
starting with $\pi$ and $a_1$ in large $N_c$.  The above-mentioned
constraints lead to relations between these resonance parameters,
ChPT parameters and condensates.

Since eq.~(\ref{RESON}) contains infinitely many parameters,
we need additional input,
which we will take to be the LMD {\it ansatz},\cite{perisetal}
\begin{equation}
\frac{1}{\pi}{\rm Im}\,\Pi_V(t)=2f_\rho^2m_\rho^2
\delta(t-m_\rho^2)+\left(\frac{4}{3}\frac{N_c}{(4\pi)^2}+\dots
\right)\theta(t-s_0)\,,\label{LMD}
\end{equation}
where the second term is obtained from large-$N_c$
QCD perturbation theory (pQCD), starting from a scale $s_0$.  A similar
{\it ansatz} is used in the axial channel, with the same 
$s_0$, motivated by the fact that pQCD does not break chiral symmetry.
The parameter $s_0$ is a parameter to be determined.
Clearly, one needs $s_0>m_\rho^2,m_{a_1}^2$ for consistency.
More explicit resonances can be put in.

Considering $\Pi_{V,A}$ for large space-like
$Q^2$ term by term in the (large-$N_c$)
OPE, one obtains (from the first two terms) the well-known Weinberg
sum rules, and from the four-fermion condensate the new relation
(in the chiral limit, and to leading order in $\alpha_s$)\cite{golper}
\begin{equation}
1-\frac{3}{4}\frac{(4\pi)^4}{N_c^2}f_\rho^4=
-\frac{7}{11}\left(\left(1-\frac{f_\pi^2}{f_\rho^2m_\rho^2}\right)^{-1}
-\frac{3}{4}\frac{(4\pi)^4}{N_c^2}f_\rho^4\right)\,,\label{SE}
\end{equation}
\hrule  width5cm
\noindent
{\footnotesize {}$^1$presenter at conference}
\vfill
\eject
\noindent
the ``7/11" rule.
{}From the first Weinberg sum rule, $f_\rho^2m_\rho^2>f_\pi^2$,
from which it follows that
$f_\rho>0.15$, just from the fact that the rhs of eq.~(\ref{SE}) has
to be negative. 
If we use also the relation\cite%
{swisscat} $f_\rho^2m_\rho^2=2f_\pi^2$,
we find, using eq.~(\ref{SE}) and the Weinberg sum rules, that
$m_\rho/f_\pi=8.8$, $m_{a_1}/m_\rho=1.4$, and $f_\rho=2f_{a_1}=0.16$,
in remarkable agreement with
experiment.  We find that $\sqrt{s_0}=4\pi f_\pi>m_{a_1}$,
hence it is not unreasonable to use pQCD above $\sqrt{s_0}$.
There are perturbative corrections to the
7/11 rule, unlike the 
Weinberg sum rules but, to the order we are working, the rule does not depend
on any condensates. Furthermore the phenomenological
results are robust to $O(\alpha_s)$ corrections.

For completeness, we also considered two-point functions for the scalar
and pseudo-scalar densities, to the same order in the OPE.
The idea is to use the gluon and four-fermion condensates predicted
in this approach in the (axial-)vector sector, and see what we
learn in the (pseudo-)scalar sector.  In this case, there
are two instead of one subtractions for each two-point function,
because of the lack of current conservation, and we end up with
only four instead of six equations, with five parameters,
the scalar mass $m_S$, $m_{\pi'}$, the amplitudes $c_m$, $d_m$
replacing $f_{\rho,a_1}$, and $\hat{s}_0$, equivalent, but not
equal, to $s_0$.
We supplemented the four equations with $L_8=c_m^2/2m_S^2
-d_m^2/2m_{\pi'}^2=0.0009(3)$.\cite{swisscat}

Our findings are as follows.  With the condensates from the vector
sector, we find no solutions with $\sqrt{\hat{s}_0}>m_{\pi'}$,  
essentially because of a clash in the equations for $\hat{s}_0$
involving dimension-six operators.
Instead, we eliminated this quantity ({\it i.e.} the terms
originating from the pQCD part of our {\it ansatz} for the
(pseudo-)scalar two-point functions) from the equations involving
the four-fermion condensate,
took $m_{\pi'}=1300$~MeV as input, and found
$c_m=41$~MeV, $d_m=27$~MeV, $m_S=860$~MeV, $\sqrt{\hat{s}_0}=2.2$~GeV.
This solution is very insensitive to the gluon condensate, $\alpha_s$,
and $m_{\pi'}$, and $700<m_S<1200$ (MeV) for $12>10^4L_8>6$.
Hence, for the central value of $L_8$,
phenomenological results are rather insensitive to
our input, and one always finds a scalar mass of order
$800-900$~MeV in the chiral and large-$N_c$ limits.  
Our result for $c_m$ is in good agreement with other estimates.%
\cite{pichetal}  Using our results as additional input in the analysis
of Peris {\it et al.},\cite{perisetal} good predictions for 
$L_{1,2,3,5,8,9,10}$
are obtained if a Nambu--Jona-Lasinio-type model 
is assumed.  We conclude with
noting that the fact that we do not find a ``perfect" solution for the
{\it combined} 
vector and scalar sectors suggests that other resonances may have to
be taken into account.

\end{document}